# Electron beam shaping: how to control the e-beam propagation along atomic columns

E Rotunno[1], A. Tavabi[2], E. Yucelen[3], S. Frabboni[4], R E. Dunin Borkowski[2], E. Karimi[5], B. J. McMorran[6], V Grillo[1*]

1. CNR-NANO, via G Campi 213/a, I-41125 Modena, Italy
2. Ernst Ruska-Centre for Microscopy and Spectroscopy with Electrons, Forschungszentrum Jülich, Jülich 52425, Germany
3. FEI Company, Europe NanoPort, Achtseweg Noord 5, 5651 GG Eindhoven, The Netherlands
4. Dipartimento di Fisica Informatica e Matematica, Università di Modena e Reggio Emilia, via G Campi 213/a, I-41125 Modena, Italy
5. Department of Physics, University of Ottawa, 150 Louis Pasteur, Ottawa, Ontario K1N 6N5, Canada
6. Department of Physics, University of Oregon, Eugene, Oregon, USA

*vincenzo.grillo@nano.cnr.it

**Abstract**

In this work we report on a detailed analysis of the propagation of high energy electron beams having different shapes in a model system, namely [100] oriented zincblende GaN crystal. The analyses are based on the comparison between a reformulated Bloch wave and multislice simulations and mainly focus on Bessel beams. In fact, considering the simplicity of the Bessel beam momentum spectrum and the symmetry of the material it is possible in some cases to give a simple description of the propagation and explain it on the bases of the free space properties of each beam. This analysis permits a deeper understanding of the channeling phenomena and of the probe intensity oscillation along the propagation direction. For comparison we will also consider two additional relevant cases of the well-known aperture limited beams and a newly introduced Gaussian probes. The latter can be shown to be the optimal probe for coupling to 1s Bloch states and obtain minimal spread along columns.

# I. INTRODUCTION

Most of the beams used so far in electron microscopy are normally shaped by a hard aperture whose radius is selected in order to limit the aberrations effect [1][2]. In "normal conditions" the beam shape is close to an Airy disc. Its shape looks like a Gaussian with additional ripples which are typically visible only when the spatial coherence is very good (more ripple appears also for "unconventional" defocus or aperture size setting) [3]. However the introduction of electron vortex beams and holographic electron beam shaping has completely changed the paradigm in this field [4][5]. Probes with different kinds of complex wave fronts can be engineered. In particular vortex beams are characterized by a staircase wavefront and by a singularity in the center that produces an intensity zero exactly in the center of the beam. Nevertheless these kind of probes can still be used for the acquisition of atomic resolution images [6]. In spite of the exotic shapes, the radial profile of these vortex probes is still determined by a simple hard cutoff. Recently the use of more complex nanofabrication schemes has also allowed for the introduction of holographic masks manipulating amplitude and phase [7][8][9][10][11]. This has extended the range of possible beam shapes that can be engineered. In particular Bessel beams are among the most promising for practical applications. These beams are the Fourier transform of a narrow ring in the aperture plane and have seen important predecessor in the Hollow cone illumination [12][13][14][15][16]. However if very narrow distributions in radial momentum or non-vanishing topological charge is to be achieved, nothing beats the holographic approach [8].

Bessel beams already appear to be very promising for high resolution imaging and tomography but higher order Bessel beams are also interesting as alternative shapes for Vortex beams. They also represent a new way to manipulate the overall radial function with important consequences on the resolution. Indeed, Bessel beams show the narrowest central peak when compared with conventional beams having the same maximum convergence [17].

The large interest in Bessel beams is also motivated by the fact that these beams are propagation invariant in vacuum and are "self healing", namely insensitive to a partial obstruction by opaque objects. The natural question that may arise is if these beams are also insensitive to the propagation inside a crystal that can mainly be considered as a phase and amplitude object. Nevertheless the Bessel beam propagation in vacuum and the channeling in a material present interesting similarities as both are solutions of the paraxial Schrodinger equation obtained by variable separation between in-plane and out-of plane z component. We will therefore highlight the conserved feature on the Transverse Energy spectrum when passing from vacuum to channeling solutions.

Moreover we will try to reformulate the Bloch wave simulation paradigm based on the concept of Transverse Energy as the only quantum number. The approach has the advantage of being less related to the probe decomposition in plane waves allowing the new eigenstates to directly match the overall probe shape

At this point it is worth pointing out that the Bloch wave analysis for aperture limited vortex beams has been already carried out with interesting results [18][19] but Bessel beams, at any order, present an intriguing aspect that is worth investigating: they are characterized by a single value (or a narrow distribution) of the modulus of the transverse momentum. In high symmetry conditions (the probe sitting exactly on the column, large separation between columns and small effects from light atoms) the influence of the azimuthal coordinate is also small and this simplifies the treatment of the Bloch wave propagation.

In this work we will consider in particular the case of zincblende GaN observed along its [100] zone axis owing to its simple symmetry, despite this allotropic form of GaN not being the thermodynamically favored one. The small potential due to the N atoms can be neglected and the main potential is due to well separated Ga atoms. Therefore we can assume, to a good degree of approximation, that the Ga columns produce a azimuthally symmetric potential that simplifies the treatment. In this work we aim to develop the entire formalism for this simplified case and to consider the special case of the 0-th order Bessel beam, The zeroth order case, compared to ordinary probe, is particularly didactic for its simplicity and it provides a way to understand in detail the "pendellösung" oscillation with consequences of general interest, such as the well-known damping effect of the oscillation, with the interesting result that the damping is not due to inelastic effects.

Coupling this aspect with the discrete momentum spectrum of the Bessel beams produces, as a result, a very strong selection on excited Bloch states. We then foresee the possibility to engineer the pendellösung oscillation of the probe in STEM experiments.

Finally, for the sake of comparison, we will introduce the actual diffraction free solution of the propagation of beams in a crystal to highlight the difference with the Bessel beam. In fact, shaping the beam as an approximate 1s state allows the minimization of the diffraction/pendellösung effects observed in both aperture limited and Bessel probes. The wide variety of beam behaviors with thickness and the possibility of control given by advanced beam shaping will be made clear.

## II. GENERALITIES ON THE BEAM PROPAGATION

While the description of the beam propagation inside a material is quite complex, we will base our discussion on a simplified approach.

In general, the beam entering the sample can be decomposed in a sum over different Bloch waves $b_{\bar{k}}^n(\bar{r}, z) = b_{\bar{k}}^n(\bar{r}) \exp\left(i K_z^{(n,\bar{k})} z\right)$ described by quantum numbers n, and $\bar{k}$. Here, n indicates the band and $\bar{k}$ is the 2D pseudo-momentum confined to the first Brillouin zone, similar to the solid state description of electrons. [20]

In the case of a convergent STEM probe each partial plane wave inside the illumination cone will give rise to its own set of Bloch waves and the overall wavefunction can be expressed as [21]:

$$\Psi(\bar{r}, z) = \int \sum_n \varepsilon^{n,\bar{k}} A(\bar{K}_\perp) b_{\bar{k}}^n(\bar{r}) \exp\left(i K_z^{(n,\bar{k})} z\right) d\bar{K}_\perp$$

where $\bar{K}_\perp$ is the transverse component of the incident beam momentum, $A(\bar{K}_\perp)$ contains the aperture effect (such as any complex amplitude modulation of the probe-forming hologram) and the lens aberrations, and $\varepsilon^{n,\bar{k}}$ is the complex excitation for each Bloch state $b_{\bar{k}}^n(\bar{r}, z)$.

For many practical purposes we can simplify the description of an electron beam located on an atomic column to the two most important components:

1) $B_{\bar{k}}^{TB}(\bar{r}, z)$ single column localized states that are therefore treated as independent from the presence of other columns. This is equivalent to the "tight binding approximation" in solids. The paradigmatic example are the 1s ground states that we can call $B_{\bar{k}}^{1s}(\bar{r}, z)$

2) $B_{\bar{k}}^{HE}(\bar{r}, z)$ asymptotically free states whose propagation is approximately independent from the presence of the lattice potential.

Such components can be mathematically described as the superposition of Bloch states. In the case of the bound states the "tight binding" assumption permits the description of each state with band state n=1 and different $\bar{k}$ as $b_{\bar{k}}^{1s}(r,z) \approx b^{1s}(r)\exp(i\bar{k}\bar{r})\exp(iK_z^{1s}z)$.

The $K_z^{(n)}$, namely the velocity of phase evolution along z, can be calculated by diagonalizing the Schrödinger equation on the Bloch basis. Depending on the formalism chosen, $K_z$ can be also related to the transverse energy

$$E_T = \frac{\hbar^2}{2m}\left(K^2 - K_z^{(n)2}\right), \tag{1}$$

where $K$ is the total momentum of the electrons. For simplicity reasons, the transverse energy is expressed as $\left(K_z^{(n)2} - K_z^2\right)$ following the formalism used by Metherel in [22], i.e. we put all the constant in eq.1 equal to 1. In this way, all the bound states have a positive transverse energy, while unbound states are characterized by negative values.

$K_z^{(n)}$ can also be related to the "anpassung" parameter that is also the eigenvalue of the simplified Bloch diagonalization problem

$$\gamma = \frac{\eta}{\lambda} - K_z^{(n)}, \tag{2}$$

Here E is the beam energy after subtraction of the mean inner potential contribution, $\lambda$ is the electron wavelength in vacuum, $\eta$ is the correction to the wavelength due to mean inner potential, $m$ is the electron relativistic mass. Based on these relationships we will refer equivalently to $K_z^{(n)}$, anpassung or transverse energy.

The overall tightly bound wavefunction component for the 1s state is:

$$B_{\bar{k}}^{1s}(\bar{r},z) = \int \varepsilon^{1s,\bar{k}} A(\bar{K}_\perp) b_{\bar{k}}^{1s}(\bar{r},z) \exp\left(iK_z^{(i,\bar{k})}z\right) d\bar{K}_\perp \tag{3}$$

$$\approx \exp(iK_z^{1s}z) \int \varepsilon^{1s,\bar{k}} A(\bar{K}_\perp) b^{1s}(\bar{r}) \exp(i\bar{k}\bar{r}) d\bar{K}_\perp$$

The in-plane description of the Bloch state as $b^{1s}(\bar{r})\exp(i\bar{k}\cdot\bar{r})$ implies that there exists a single (typically Gaussian shaped) mode $b^{1s}(\bar{r})$ independent of $\bar{k}$ and the $\bar{k}$ dependence on the full Bloch wave $b_{\bar{k}}^{1s}(r,z)$ is confined to the phase factor $\exp(i\bar{k}r)$. This assumption is very similar to what happens in the k.P model in solid state theory of Bloch electrons.

In this case, as we consider nearly isolated Ga columns, we will have a small azimuthal dependence of the excitation factors $\varepsilon^{1s,\bar{k}}$ with a further simplification in the data interpretation.

As for the $B_{\bar{k}}^{HE}(\bar{r},z)$ the description is also simple and we can write

$$B_{\bar{k}}^{HE}(\bar{r},z) = \int \varepsilon^{HE,\bar{k}} A(\bar{K}_\perp) \exp(iK_z^{HE}\bar{r}) d\bar{K}_\perp \tag{4}$$

the relation between $\overline{K}_\perp$ and $K_z^{HE}$ can be written

$$K_z^{HE} = \sqrt{\frac{\eta^2}{\lambda^2} - K_\perp^2} \approx \frac{\eta}{\lambda}\left(1 - \frac{1}{2}\frac{\lambda^2}{\eta^2}K_\perp^2\right) \tag{5}$$

In this case we do not write the relation to the pseudo momentum $\overline{k}$ as we assume that we can neglect the crystalline potential with respect to transverse energy of these states.

This very simplified description can be employed to qualitatively analyze different kinds of beam propagations, as consequence of the beating between these two components.

In order to study in detail the propagation, we analyzed the traverse energy spectrum of the excited states. Rather than using the Bloch states quantum number *n,k* where only n is the discrete variable, we labeled states based on their transverse energy only. Namely states with different n,k number are grouped together depending on their transverse energy $\tilde{E}$ to form a new state:

$$B^{\tilde{E}}(\bar{r},z) = \exp(iK_z^{\tilde{E}}z) \int \sum_n \varepsilon^{n,\overline{k}} A(\overline{K}_\perp) b_{\overline{k}}^n(\bar{r},z) \delta(E^{n,\overline{k}},\tilde{E}) d\overline{K}_\perp \tag{6}$$

The propagation inside the crystal at a point $\bar{r}$ can be seen as the effect of the z dependent interference of the states $B^{\tilde{E}}(\bar{r},z)$ with intensity $\left|B^{\tilde{E}}(\bar{r},z)\right|^2$ but for many purposes we consider a unit cell averaged intensity $I(\tilde{E}) = \int \left|B^{\tilde{E}}(\bar{r},z)\right|^2 d\bar{r}$.

Since the Bloch wave algorithms produce the Bloch wave parameters and excitation for plane wave only, we sampled a number of points within the probe, depending on the probe geometry, and summed the results together accounting for the appropriate phase term, namely the aberrations phase. However for simplicity reasons, we consider here the case of a perfectly aberration-corrected microscope in which all residual aberrations have been set to zero. The actual intensity spectrum has been produced using several times the EMS code (bz function) [23] controlled from inside STEMCELL.

Studying the intensity spectrum as a function of the state's energy gives a straightforward way to understand the propagation of any beam. TB states will appear on the spectrum as sharp peaks while HE states will form wide bands which average value will give the mean velocity of the group as it will be explained in the following discussion.

It is also possible to explain the pendellösung by evaluating the dephasing of the different Bloch waves. Considering a simple distribution $P(\Delta k)$ of continuum states, the wavefunction amplitude in a single (x,y) point can be expressed as:

$$\psi(z) = \frac{1}{N}\int_{-\infty}^{+\infty}\left[P(\Delta k)e^{i\Delta k\, z}e^{ik\, z} - e^{ik_0\, z}\right]d\Delta k \tag{7}$$

This means that the actual wavefunction is:

$$\psi(z) = \check{P}(z)e^{ikz} + e^{ik_0 z} \tag{8}$$

Where $\check{P}(z)$ is the Fourier transform of the distribution $P(\Delta k)$.

That gives an intensity:

$$I(z) = |\check{P}(z)|^2 + 1 + 2\check{P}(z)\sin((k-k_0)z) \tag{9}$$

Where $\check{P}(z)$ describes the damping profile of the pendellösung oscillation as a function of the depth coordinate. This is a noticeable result as the damping of the oscillation is not due to inelastic effects, as commonly thought.

This means that if the momenta along z are distributed, for example, according to a Gaussian with size δk, then the oscillation is damped within a distance $\delta z = 2\pi/\delta k$.

### III. RESULTS

### A. Conventional aperture-limited probe

The description of a conventional probe (aperture limited) is straightforward: since the states HE are nearly free states they behave as the probe would do in vacuum; the $B_K^{HE}(\bar{r}, z)$ forms a concentrated waist inside the sample at a depth corresponding to the in-focus condition and then tends to broaden, for geometric reasons, to a radius R=zα, where α is the convergence. In an alternative language we can say that the component HE states get out of phase before and after the focus.

Since the $B_K^{1s}(\bar{r}, z)$ overall phase $\exp(iK_z z)$ evolves with z faster than that of all the $B_K^{HE}(\bar{r}, z)$ states, the two states give rise to characteristic beatings with frequency

$$K_z^{1s} - K_{0_z}^{HE} \tag{10}$$

Where $K_{0_z}^{HE}$ is the average propagation velocity over all the HE states.

However, since the $B_K^{HE}(r, z)$ states gets quickly out of phase with each other, the oscillation is rapidly damped. This point is worth some analyses since it is commonly believed [24][25] that the oscillations damping is due to the absorption of the 1s states that instead here we intentionally neglect.

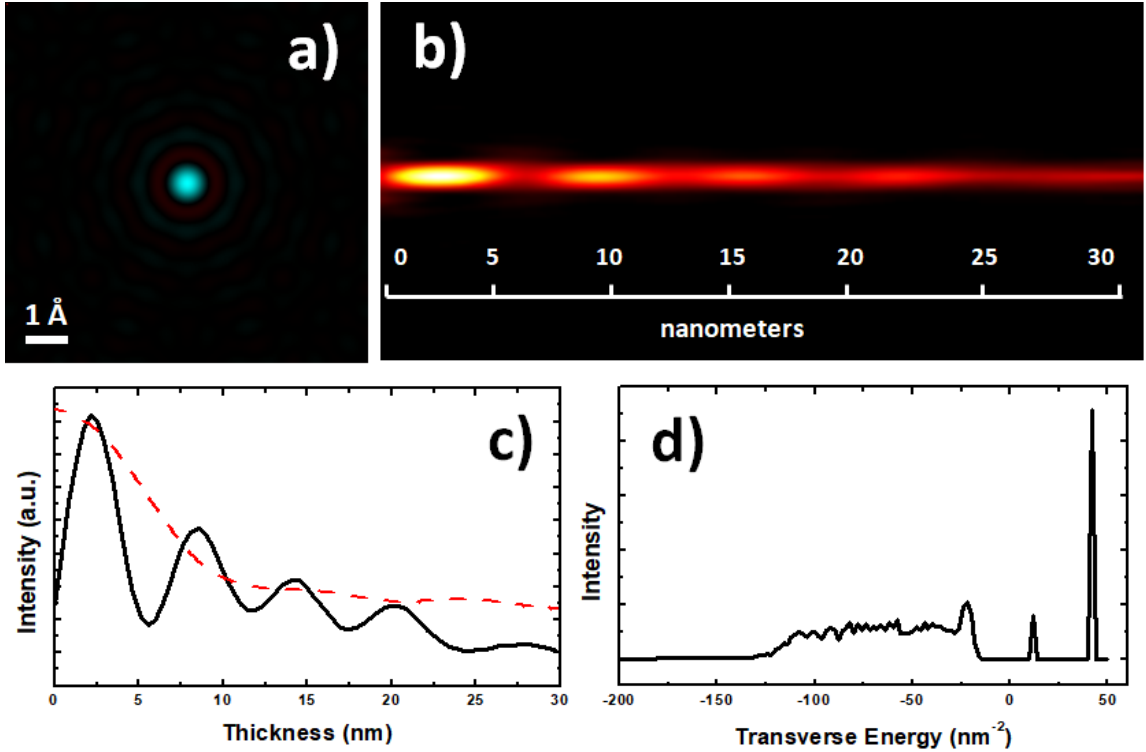

**Figure1: a)** simulated image of a probe formed by a 21 mrad aperture at 300KeV and its evolution **(b)** along a Ga column in a [100] oriented GaN crystal. **c)** intensity line profile as function of the depth (solid black line) along with an estimated damping profile P(z) (red dashed line). **d)** transverse energy spectrum of the excited Bloch states intensities.

In figure 1 the result of the multislice simulation is reported. It has been performed using a routine inside the STEMCELL software suite [26], derived by Kirkland multislice code. The results are obtained for a [100] zincblende GaN column for a probe formed by a 21 mrad aperture at 300KeV (Figure 1a) located on the Ga column. The probe intensity along the column is reported in figure 1b. The characteristic channeling oscillation are clearly visible as highlighted by the representation of the intensity as function of the depth plotted in figure 1c.

Finally, the traverse energy spectrum of the excited Bloch states intensities has been calculated by sampling the probe into 1793 reciprocal points, and the result is shown in figure 1d.

The traverse energy spectrum of the excited Bloch states intensities is dominated by a broad band roughly extending from -10 nm$^{-2}$ to -130 nm$^{-2}$ that corresponds to the eigenvalues of $B_K^{HE}(\bar{r},z)$ states. In the positive transverse energy part of the spectrum a very sharp peak is present at about 45 nm$^{-2}$; it is ascribed to the tightly bound, non dispersive 1s state. Moreover, the 1s state intensity is about 1 order of magnitude higher than any other Bloch state, as expected in the case of strong channeling in a crystalline solid oriented along a major zone axis.

The characteristic channeling oscillations then arise as a direct consequence of the interference between the 1s Bloch states with each one of the $B_K^{HE}(\bar{r},z)$ states. Their frequency can be readily quantified from the intensities spectrum: the average transverse energy of the $B_K^{HE}(\bar{r},z)$ states is about -80 nm$^{-2}$ which can be directly translated in term of group velocity using equation 2; equation 10 finally gives the frequency of the beating. The periodicity is estimated in ≈8 nm in very good agreement with the multislice calculation.

Clearly, due to the large spread of propagation velocity, the $B_K^{HE}(\bar{r}, z)$ wave packet rapidly disperses and a quick damping of the oscillation can be expected, as described in equation 9. Of course for a probe with larger convergence, the band is broader and a larger damping is produced as expected by the fact that the $B_K^{HE}(\bar{r}, z)$ are spread out at a faster rate.

The calculated damping profile is reported as the red dashed line in figure 1c and it is in good agreement with the multislice calculations.

### B. Bessel probe

The general form of the time independent Bessel beam solution of order n is simply

$$\Psi(\rho, \phi, z; t) = J_n(k_\rho, \rho) e^{in\phi} e^{i(k_z \cdot z)} \tag{11}$$

where $J_n$ represents an n-th order Bessel function of the first kind, n is an integer, $k_\rho$ and $k_z$ are respectively the wavefunction's transverse and longitudinal wave vector components, is related to its de Broglie wavelength $\lambda$ by the relation $k^2 = k_\rho^2 + k_z^2 = \frac{2m\omega}{\hbar} = \left(\frac{2\pi}{\lambda}\right)^2$, where k is the modulus of the electron wavevector and $\hbar$ is the reduced Planck constant. Here we will consider only the case of n=0.

Moreover, Bessel probes (like plane waves) have non-normalizable intensity and can be only approximated by truncated Bessel beam that correspond to a finite annulus size in hollow cone illumination. Here, the truncation has been appropriately chosen in order to have most of the wavefunction intensity within a single unit cell (Figure 2a).

Even in these approximated conditions, a Bessel probe propagating in the vacuum has a very long range of defocus for which it remains localized [27]. By similarity with the vacuum case, we can also assume that the $B_K^{HE}(\vec{r}, z)$ component of the beam does not spread along Z, as it will be discussed in details later.

Multislice calculations of the Bessel beam propagation are reported in figure 2b and c. They show a much clearer oscillation with respect to the conventional probe case (figure 1). In fact, nearly no damping of the oscillation is visible in the first 30nm of the propagation. It is also worth noticing that the oscillation frequency is similar, with spurious differences due to the change of the barycenter in the HE states distribution in the two cases.

Consistently, the intensities spectrum, figure 2d, calculated by sampling the probe into 508 reciprocal points, contains a relatively narrow peak in the negative transverse energy regime at around -115 nm$^{-2}$. In the case of an isolated column the spectrum should be perfectly monochromatic, however the presence of a small breaking of the azimuthal symmetry due to the N atomic columns, leads to a small spreading of the transverse energy. Nevertheless, the approximate azimuthal symmetry allows for a clear interpretation of the spectrum that would not work for more complicated symmetries or just for small probe misplacement.

Finally, the damping profiles, calculated according equation 9, is reported in figure 2c (red dashed line). The simple shape of the HE states band, that can be approximated to a Gaussian distribution, ensures a nearly perfect quantitative agreement with the multislice calculations.

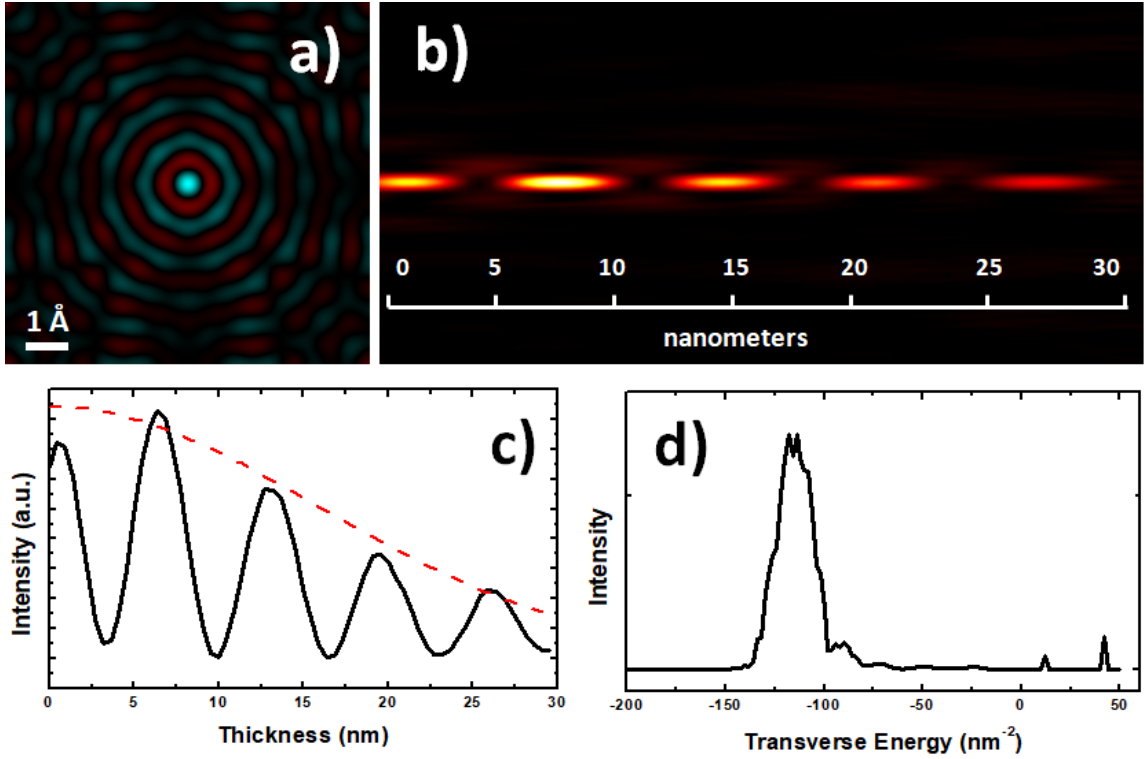

**Figure 2: a)** simulated image of a Bessel probe formed by a 20-22 mrad ring aperture at 300KeV and its evolution **(b)**; along a Ga column in a [100] oriented GaN crystal. **c)** intensity line profile as function of the depth. The red dashed curve represent the damping function calculated according appendix 1. **d)** transverse energy spectrum of the excited Bloch states intensities.

Both these arguments can be claimed to explain the results of the multislice calculations. In general, we can say that the behavior of a Bessel probe along an atomic column in a crystal is largely similar to the in vacuum propagation with the addition of the beating with the highly localized 1s states. The persistence of the oscillation is therefore a consequence of the non-diffractive nature of Bessel beams in vacuum and, for extension, of unbound states in crystals.

Considering the real space shape of the $B_K^{HE}(\vec{r},z)$ and $B_K^{1s}(\vec{r},z)$ components, an interesting detail appears: $B_K^{HE}(\vec{r},z)$ is clearly a Bessel function, namely $B_K^{HE}(\vec{r},z) = J_0(k_{\rho HE},\vec{r})\exp(iK_z^{HE}z)$ but also $B_K^{1s}(\vec{r},z)$ can be approximated, from equation 4, to be $B_K^{1s}(\vec{r},z) = b^{1s}(\vec{r})J_0(k_{\rho 1s}\vec{r})\exp(iK_z^{1s}z)$ that is a Bessel-Gauss function.

We have therefore the noticeable results that the whole beam is the superposition of two beating Bessel functions.

In multislice simulations we can easily highlight this by calculating the probe intensity at the oscillation maxima, as demonstrated in figures 3 a) and b) showing the formation of two sets of characterizing rings with different radii.

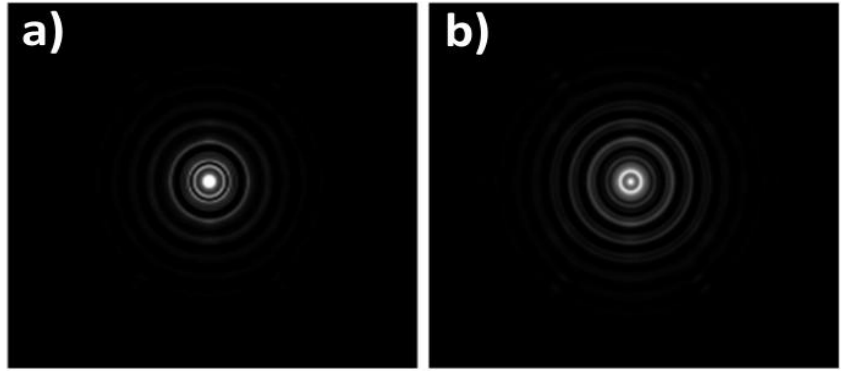

**Figure 3:** Simulated image of a Bessel probe formed by a 20-22 mrad ring during its evolution inside the specimen, at (a) 6 and (b) 12 nm depth respectively.

## IV. DISCUSSION
### A. Engineering the Channeling

Having proved these properties of Bessel beam we can try to work out the dependence on a few parameters. In particular, in figure 4 we report the multislice simulations of the propagation of Bessel beams having a) 20-22 mrad, c) 16-18 mrad and e) 14-16 mrad convergences. It can be observed that, by varying the convergence semi angle it is possible to vary the frequency of the beating along the atomic column. Conversely, if we change the defocus at the entrance of the sample we don't observe any shift of the beating fringes.

As an example, in figure 4g, we applied a 10 nm defocus to the 20-22 mrad probe. The effect of the applied defocus is almost negligible as shown by the intensity line profile of figure 4h where the profile of the probe without defocus is overlapped in red. This behavior is completely explained by the discussion above and the properties of Bessel beam in vacuum.

We can make an intelligent use of this behavior connecting to a precedent paper.[28] In fact we already demonstrated that probes with different channeling behavior can be jointly used to produce 3D information about guest atomic species in a lattice. In fact, the contribution of an atom, located at a depth z, to the total image intensity is the product of the probe current in that position, j(z), times the atomic scattering cross-section.

Therefore, the image contrast can be related to the channeling current j(z) and the distribution of guest species in the column a(z) through the simple relation:

$$C = \sum_{i\ atoms} j(z)a(z) \quad (12)$$

In the specific case of Bessel beams the channeling current j(z) are just trigonometric functions. The imaging with different convergences produces a harmonic decomposition of the unknown aimed function a(z).

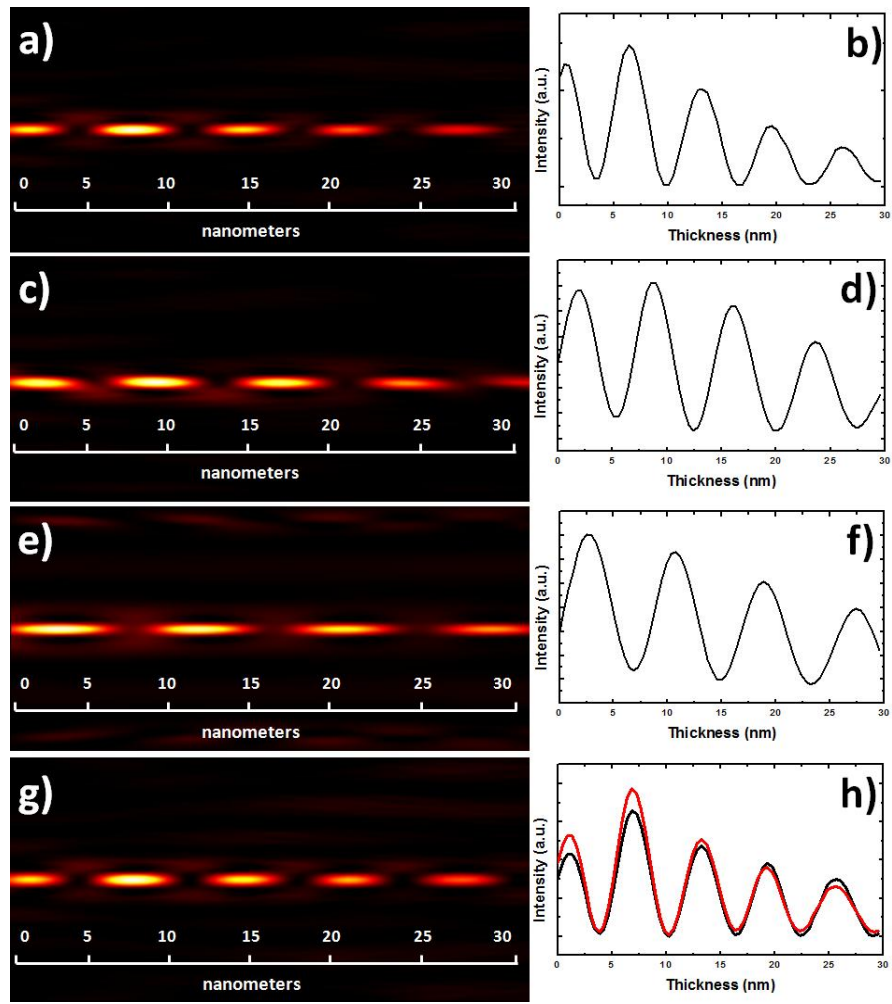

**Figure 4: a)** multislice simulation of the propagation along a Ga column in a [100] oriented GaN crystal of a Bessel probe formed by a **a)** 20-22 mrad ring aperture and **b)** its intensity line profile as function of the depth; **c)** 16-18 mrad ring aperture and **d)** its intensity line profile as function of the depth; **e)** 14-16 mrad ring aperture and **f)** its intensity line profile as function of the depth; **g)** 20-22 mrad ring aperture with 10 nm defocus and **h)** its intensity line profile as function of the depth. The profile of figure 4b) is overlapped in red as a reference.

### B. The optimal channeling probe

As a further step, based on the above considerations, we could wonder what would be the ideal probe having the same role in the material that Bessel beams have in vacuum, i.e. that propagates in the crystal without diffracting. This problem is equivalent to ask what are the solutions of the wave equation inside the material. Fortunately, the answer is well known in microscopy and these are just the 2D-Bloch waves solution.

Whereas it is complicated to produce a single Bloch waves of arbitrary order that would be completely delocalized it is in principle possible to produce a beam resembling the 1s states by beam synthesis techniques. To a very large approximation it can be considered as a Gaussian beam (Figure 5a). In this specific case, the beam intensity has a FWHM =0.35Å namely close to the current instrumental limits.

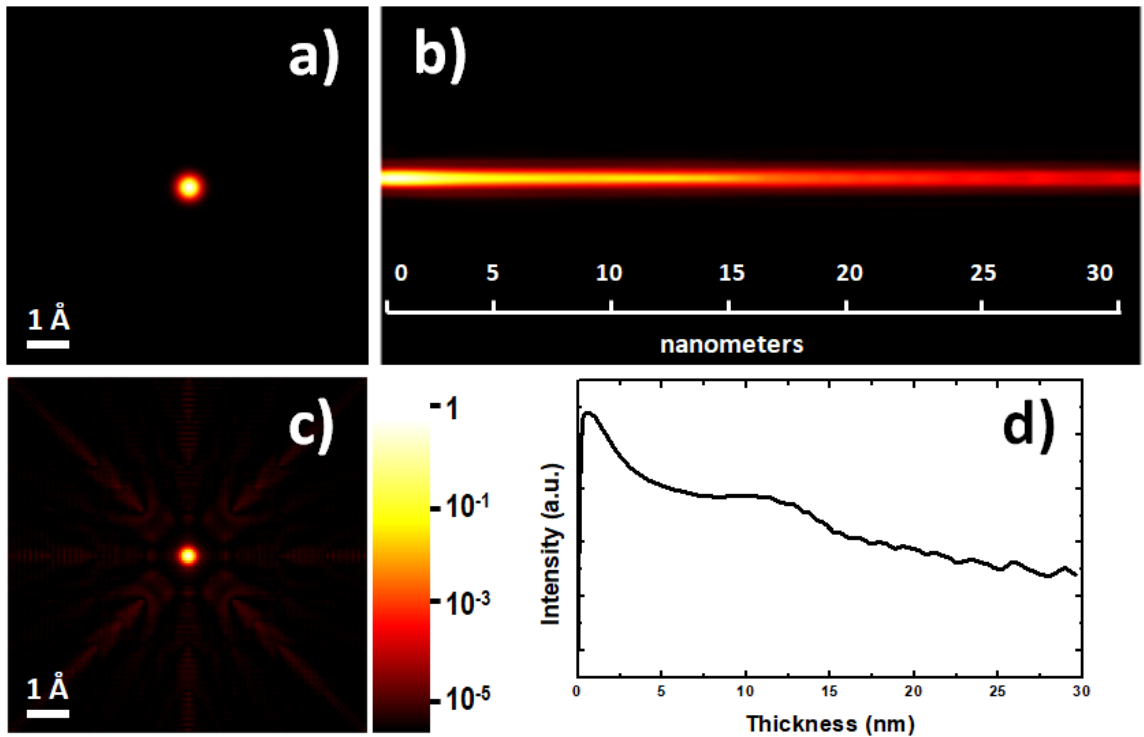

**Figure 5: a)** simulated image of a Gaussian probe at 300KeV and its evolution **(b)** along a Ga column in a [100] oriented GaN crystal. **c)** simulated image of the Gaussian probe at the exit surface after the propagation inside the specimen. **d)** intensity line profile as function of the depth.

Figure 5b shows the simulation for the propagation of such a probe inside the GaN cell as above, while its depth dependence is shown in figure 5d. The Gaussian probe is still affected by the absorption effects but practically no sign of beating between states is visible.

Finally, figure 5c shows, in logarithmic scale, the probe at the end of the propagation. Clearly the probe shows very week signs of cross talk between adjacent columns.

These results, along with the previous considerations, open up a completely new field in the engineering of the probe with a very wide field of applications that will be considered in further coming articles.

## V. CONCLUSIONS

In this work we have coupled together Bloch wave and Multislice simulations to comprehensively study the behavior of Bessel and optimal Gaussian beams inside a material.

We studied in particular the case of a nearly isolated Ga column in GaN [100] with the probe localized on Ga atoms. In the case for even states the probe can be seen as the superposition of few relevant states. So we found that for 0-th order Bessel probe the probe can be represented as superposition of two Bessel beams with different z velocities.

The case of Bessel beam has been compared with normal "aperture-limited" probes permitting to understand the dynamic of the "pendellösung". The comparison allowed us to clarify that the damping of channeling is due to the dispersion of the non-bound states rather than absorption of the 1s states.

Finally we briefly compare the results with the theoretically optimal probe for channeling, namely a Gaussian beam with the same size of the 1s Bloch wave of the material. Not surprisingly this is the only probe showing no oscillation in the channeling behavior and it is the right complement of Bessel probe.

**Acknowledgements**

The work is supported by Q-SORT, a project funded the European Union's Horizon 2020 Research and Innovation Programme under grant agreement No.766970. Work at University of Oregon was supported by both the U.S. Department of Energy, Office of Science, Basic Energy Sciences, under Award DE-SC0010466.

**References**

[1] O. Scherzer, J. Appl. Phys. **20,** 20 (1949).

[2] V Intaraprasonk , H L. Xin , D A. Muller, Ultramicroscopy **108,** 1454– 1466 (2008).

[3] K. Watanabe, N. Nakanishi, T. Yamazaki , M. Kawasaki, I. Hashimoto and M. Shiojiri, Phys. Stat. Sol. (b) **235**, 179–188 (2003).

[4] J. Verbeeck, H. Tian, P. Schattschneider, Nature **467**, 301–304 (2010).

[5] B. J. McMorran, A. Agrawal, I. M. Anderson, A. A. Herzing, H. J. Lezec, J. J. McClelland, J. Unguris, Science **331**, 192-195 (2011).

[6] J. Verbeeck, H. Tian, A. Béché, Ultramicroscopy **113**, 83–87 (2012).

[7] V. Grillo, G. C.o Gazzadi, E. Karimi, E.Mafakheri, R. W. Boyd, and S. Frabboni, Appl. Phy. Lett. **104**, 043109 (2014).

[8] V. Grillo,E. Karimi, G. C. Gazzadi, S. Frabboni, M. R. Dennis and R. W. Phys. Rev. X **4**, 011013 (2014).

[9] V. Grillo, E. Rotunno, B. McMorran, S. Frabboni, Microsc. Microanal. 1889-1890, (2015).

[10] T. R. Harvey, J. S. Pierce, A. K. Agrawal, P. Ercius, M. Linck, and B. J. McMorran, New J. Phys. **16**, 093039 (2014).

[11] B. J. McMorran, T. R. Harvey, J. Perry-Houts, S. Cabrini, A. Agrawal, and H. Lezec. EIPBN Proceedings (2012) P04-16.

[12] C. Zheng, T. C. Peterson, H. Kirmse, W. Neumann, M. J. Morgan, J. Etheridge, Phys. Rev. Lett. **119**, 174801 (2017).

[13] W.D. Riecke, Z. Naturforschg. **19a,** 1228 (1964).

[14] J. M. Gibson, and A. Howie, Chem. Scripta **14,** 109–116 (1979).

[15] T. Kawasaki , T. Matsutani , T. Ikuta , M. Ichihashi , T. Tanji, Ultramicroscopy **110,** 1332–1337 (2010).

[16] D.T. Nguyen, S.D. Findlay, J. Etheridge, Ultramicroscopy **184**, 143-155 (2018).

[17] J. Durnin, J. J. Miceli, J. H. Eberly. Opt. Lett. **13,** 79 (1988).

[18] B.G. Mendis, Ultramicroscopy **149**, 74–85 (2015).

[19] H.L. Xin and H. Zheng, Microsc. Microanal. **18**, 711–719 (2012).

[20] Hirsch, PB, Howie, A, Nicholson, RB, Pashley, DW and Whelan, MJ 1977 Electron Microscopy of Thin Crystals 2nd Ed. Krieger Huntington NY.

[21] Y. Peng, P. D. Nellist, S. J. Pennycook, Journal of Electron Microscopy **53** 257–266 (2004).

[22] A. J. F. Metherell in Electron Microscopy in Materials Science II (1975) 397–552 Eds. U. Valdre´ and E. Ruedl, CEC, Brussels.

[23] P.A. Stadelmann, Ultramicroscopy **21,** 131-145 (1987).

[24] P.M. Voyles D.A. Muller .E.J. Kirkland, Microsc. Microanal. **10,** 291–300 (2004).

[25] S.J. Pennycook C.S. Rafferty, P.D. Nellist, Microsc Microanal **6,** 343–352 (2000).

[26] V. Grillo and E. Rotunno, Ultramicroscopy **125**, 97-111, (2013)

[27] V. Grillo, J. Harris, GC Gazzadi, R. Balboni, E. Mafakheri, MR Dennis, S. Frabboni, RW Boyd, E. Karimi. Ultramicroscopy **166**, 48-60 (2016).

[28] E. Rotunno, M. Albrecht, T. Markurt, T. Remmele, V. Grillo. Ultramicroscopy **146,** 62–70 (2014).